\documentclass[preprint]{elsarticle}

\usepackage{amsmath}
\usepackage{amssymb}
\usepackage{graphicx}
\usepackage{dcolumn}
\usepackage{bm}
\usepackage{epstopdf}

\newcommand{\Ga}{\Gamma}

\newcommand{\om}{\omega}
\newcommand{\prt}{\partial}

\newcommand{\br}{\mathbf{r}}
\newcommand{\bu}{\mathbf{u}}
\newcommand{\im}{\mathrm{Im}}

\begin{document}

\title{
Condition for convective instability of dark solitons}

\author{A. M. Kamchatnov}
\ead{kamch@isan.troitsk.ru}
\author{S. V. Korneev}
\ead{svyatoslav.korneev@gmail.com}

\address{
Institute of Spectroscopy, Russian Academy of Sciences, Troitsk,
Moscow Region, 142190, Russia }

\date{\today}

\begin{abstract}
Simple derivation of the condition for the transition point from absolute instability of plane dark solitons
to their convective instability is suggested. It is shown that unstable wave packet expands with
velocity equal to the minimal group velocity of the disturbance waves propagating along a dark
soliton. The growth rate of the length of dark solitons generated by the flow of Bose-Einstein
condensate past an obstacle is estimated. Analytical theory is confirmed by the results of numerical simulations.
\end{abstract}


\maketitle

\section{Introduction}

Instability of dark solitons with respect to transverse perturbations is well studied both theoretically
and experimentally (see, e.g., review articles \cite{kivshar-98,kivshar-2000,frantz-10} and references therein).
Qualitatively, it is caused by decrease of the energy per unit length of a soliton with increase of
the local velocity of the disturbed element. As a result, the stretched segments of the soliton acquire
increased velocity and hence the soliton's local curvature grows with time which leads to
breaking of the soliton followed by formation of vortex pairs. However, this mechanism of the dark
soliton instability was studied mainly when the dark soliton propagates in a quiescent medium. For example,
this situation is realized in dynamics of Bose-Einstein condensate (BEC), when dark solitons are formed by
means of density or phase engineering. However, the situation can change drastically, if there is fast
enough flow of the condensate along the dark soliton. In particular, as was found in \cite{egk-2006},
just this happens when dark solitons are generated by a flow of BEC past an obstacle. In this case
dynamics of BEC is described (in standard non-dimensional units) by the Gross-Pitaevskii (GP) equation
\begin{equation}\label{1-1}
    i\psi_t+\tfrac12\Delta\psi-(|\psi|^2-1)\psi=U(\br)\psi,
\end{equation}
where $\psi(\br, t)$ is the condensate wave function and  $U(\br)$ the potential of the obstacle.
This equation was solved numerically in \cite{egk-2006} with the boundary condition
\begin{equation}\label{1-2}
\psi |_S = \exp(iMx)
\end{equation}
at the boundary $S$ far enough from the obstacle, and it was found that there is a critical value of
the Mach number of the incident flow,
\begin{equation}\label{1-3}
M_{cr}\cong 1.43,
\end{equation}
so that for $M<M_{cr}$ the vortices are generated downstream the obstacle whereas for $M>M_{cr}$
the dark solitons are stretched from the shadow behind the obstacle. This difference in the flow
behavior was explained in \cite{kp-2008} as a transition from absolute instability of dark solitons
to their convective instability, so that for $M>M_{cr}$ the unstable disturbances of the dark soliton
are convected by the flow along it with such a velocity that they cannot destroy the soliton at any
finite distance from the obstacle after long enough time and, hence, the soliton increases its length
with time. In this letter we shall introduce the notion of velocity of propagation of the front of
the unstable wave packet into undisturbed region of the dark soliton. This will permit us to give
simple physical interpretation of the condition for the absolute/convective instability transition found
in \cite{kp-2008} and to determine quantitatively the growth rate of oblique solitons length in the
experiments with flow of condensate past an obstacle what is quite topical in connection with recent
experimental observation \cite{amo-2011} of oblique solitons in the flow of the polariton condensate.

\section{Absolute and convective instability of dark solitons}

To simplify the notation, equation (\ref{1-1}) is written in such units that the background uniform
density is equal to unity and unit of velocity is equal to the sound speed ($c=1$). Let the dark soliton
propagate along $x$ axis with velocity $V$, then the corresponding  dark soliton solution of (\ref{1-1})
(with $U=0$) has the form
\begin{equation}\label{2-1}
\psi(x,y,t) = \psi_s(x-Vt) = \sqrt{1-V^2} \tanh [\sqrt{1-V^2}(x-Vt)] + iV.
\end{equation}
If we represent the condensate wave function as
\begin{equation}\label{2-2}
\psi(x,y,t) = \sqrt{n(x,y,t)} \exp (i \phi(x,y,t))
\end{equation}
then the condensate density and the flow velocity,
\begin{equation}\label{2-3}
n= |\psi|^2, \qquad \bu = (u, v) = \nabla \phi,
\end{equation}
are given for the soliton solution by the formulae
\begin{equation}\label{2-4}
\begin{split}
& n = n_s(x-Vt) = 1- \frac{1-V^2}{\cosh^2[\sqrt{1-V^2}(x-Vt)]} \\
& u = \frac{\prt \phi}{\prt x} = V \left (1-\frac{1}{n} \right), \quad v = \frac{\prt \phi}{\prt y} = 0.
\end{split}
\end{equation}
Stability of such a solution was studied in \cite{kt-1988}. The disturbed soliton solution can be written as
\begin{equation}\label{2-5}
\psi = \psi_s + \phi = \psi_s + \phi' + i\phi'' = \sqrt{n_s + \delta n}\, e^{i(\phi_s + \delta \phi)}
\end{equation}
where in the linear approximation the real variables $\phi'$ and $\phi''$ are related with perturbations
of the density $\delta n$ and the phase $\delta \phi$ by the formulae
\begin{equation}\label{2-6}
    \phi'=\frac{\cos\varphi_s}{2\sqrt{n_s}}\delta n+\sqrt{n_s}\,\sin\varphi_s\cdot\delta\varphi,
    \quad
    \phi^{\prime\prime}=\frac{\sin\varphi_s}{2\sqrt{n_s}}\delta n-\sqrt{n_s}\,\cos\varphi_s\cdot\delta\varphi.
\end{equation}
Substitution of (\ref{2-6}) into (\ref{1-1}) with $U(\br)=0$ and linearizing of the resulting
equation with respect to $\phi$ give
\begin{equation}\label{2-7}
    i\phi_t+\phi+\tfrac12(\phi_{xx}+\phi_{yy})-(2|\Phi_s|^2\phi+\Phi_s^2\phi^*)=0,
\end{equation}
where asterisk denotes complex conjugation. Let the disturbance be represented by a harmonic wave
propagating along the soliton: $\delta n,\, \delta \phi \propto \exp [i (py - \Omega (p) t)]$ or
$\phi=(\phi'+i\phi'')\propto\exp[i (py-\Omega (p) t)], \phi=(\phi'-i\phi'')\propto\exp[i (py-\Omega (p) t)]$.
For this harmonic form of the disturbance, equation (\ref{2-7}) reduces to
\begin{equation}\label{2-8}
    i\Omega\phi+iV\phi_{\zeta}+\phi+\tfrac12\phi_{\zeta\zeta}-\tfrac12 p^2\phi-
    (2|\Phi_s|^2\phi+\Phi_s^2\phi^*)=0,
\end{equation}
where $\zeta=x-Vt$. Substitution of (\ref{2-4}) and separation of real and imaginary parts yields the
eigenvalue problem for $\Omega=\omega(p) + i\Gamma (p)$
\begin{equation}\label{2-9}
    \left(
      \begin{array}{cc}
        L_{11} & L_{12} \\
        L_{21} & L_{22} \\
      \end{array}
    \right)
    \left(
      \begin{array}{c}
        \phi' \\
        \phi^{\prime\prime} \\
      \end{array}
    \right)=
    \left(
    \begin{array}{cc}
    \Ga & \om\\
    -\om & \Ga
    \end{array}
    \right)
    \left(
                   \begin{array}{c}
                     \phi' \\
                     \phi^{\prime\prime} \\
                   \end{array}
                 \right)
\end{equation}
where
\begin{equation}\label{2-10}
    \begin{split}
    L_{11}&=-V\prt_{\zeta}-2kV\tanh(k\zeta),\quad
    L_{12}=-\tfrac12\prt_{\zeta\zeta}-1 +n_s+2V^2+\tfrac12p^2,\\
    L_{21}&=\tfrac12\prt_{\zeta\zeta}+1-3n_s+2V^2-\tfrac12p^2,\quad
    L_{22}=-V\prt_{\zeta}+2kV\tanh(k\zeta).
    \end{split}
\end{equation}
It can be solved numerically for any given value of the soliton's velocity $V$. As was found in \cite{kt-1988},
the spectrum consist of two parts---unstable ($\Omega = i\Gamma(p)$) for $0<p<p_c$ and stable ($\Omega=\omega(p)$)
for $p>p_c$, where
\begin{equation}\label{2-11}
p_c = [-(1+V^2) + 2\sqrt{V^4-V^2+1}]^{1/2}.
\end{equation}

\begin{figure}[bt]
\begin{center}
\includegraphics[width=8cm,height=6cm,clip]{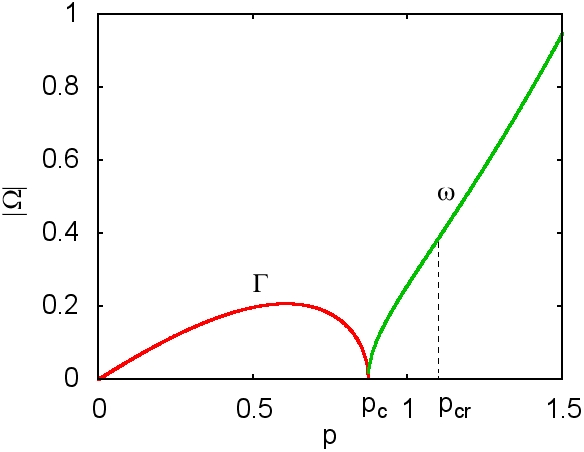}
\caption{Growth rate of unstable modes for $0<p<p_c$ (red) and dispersion law of harmonic stable
modes for $p_c<p$ (green) as functions of the wave number $p$ of perturbation.
Soliton's velocity is equal to $V=0.355$ and $p_{cr}=1.08$.
 }
\end{center}\label{fig1}
\end{figure}
Typical plots for $\Ga(p)$ and $\om(p)$ are shown in Fig.~1. Non-zero values of $\Ga$ for $0<p<p_c$ mean
instability of dark solitons.
The dispersive curve $\om(p)$ for $p_c<p<\infty$
has an inflection point $p=p_{cr}$ at which
\begin{equation}\label{2-12}
\left.\frac{d^2 \omega}{dp^2}\right|_{p=p_{cr}}  =0
\end{equation}
The found in \cite{kp-2008} criterium of transition from absolute to convective instability of oblique
solitons can be written in the form
\begin{equation}\label{2-13}
V_{\parallel {cr}} = \left. \frac{d \omega}{dp} \right|_{p=p_{cr}},
\end{equation}
where
\begin{equation}\label{2-14}
V_{\parallel_{cr}} = M_{cr} \cos \theta,
\end{equation}
is the component of the flow velocity along the soliton, $\theta$ being the angle between
the direction of the incident
flow ($x$ axis) and the oblique soliton, and the soliton's velocity $V$ is equal to
\begin{equation}\label{2-15}
V = V_{\perp} = M_{cr} \sin \theta.
\end{equation}
Hence the dispersion law $\omega = \omega(p, V)$ depends on $M_{cr}$ and (\ref{2-13}) is actually
the equation for $M_{cr}$ as a function of $\theta$. This equation was solved numerically in \cite{kp-2008}
and in the limit of small $\theta$ (deep solitons) the solution tends to the value (\ref{1-3}) from below.

\section{Velocity of expansion of unstable disturbances}

Equations (\ref{2-12}) and (\ref{2-13}) can be interpreted in general case, when the soliton's velocity
$V=V_{\perp}$ and the flow velocity $V_{\parallel}$ along it are independent parameters, as the statement
that the front of the disturbance which breaks the dark soliton with formation of vortices propagates
into undisturbed region with velocity equal to the minimal group velocity
\begin{equation}\label{2-16}
V_{cr} = \min \left(\frac{d \omega}{dp} \right).
\end{equation}
If $V_{\parallel} > V_{cr}$,
then this wave packet is convected away by the flow from a region around any fixed value of the coordinate
along the soliton and the instability is convective; if $V_{\parallel}<V_{cr}$, then the wave packet
evolves into nonlinear stage with breaking of the soliton into vortices at any fixed location along
the soliton and the instability is absolute.

This formulation can be supported by the following reasoning. At the edges of the disturbance its
amplitude is small and hence it can be described by the linear theory. Therefore we represent
evolution of a disturbance as
\begin{equation}\label{2-18}
\phi(y,t) = \int f(p) e^{i(py-\Omega(p)t)}dp
\end{equation}
For large time $t$ the integral can be evaluated by the steepest descent method so that the main
contribution to it is given by vicinity of the saddle points $p_s$ defined as solutions of the equation
\begin{equation}\label{2-19}
\frac{y}{t} = \frac{d\Omega}{dp}.
\end{equation}
The values of $p_s$ depend on $y/t$ and we suppose that for large enough $y$ equation (\ref{2-19})
yields two real solutions. Then the integral function includes the fast oscillating factor $e^{i p_s y}$
and the resulting integral corresponds to a usual dispersive wave packet slowly decaying with time as
$t^{-1/2}$. However, if with decreasing of $y$ the roots of Eq.~(\ref{2-19}) move along the real
$p$-axis and collide with each other at some value of $y=y_f$ bifurcating here into two complex roots
$p_s$ and $p^{*}_s$, then the saddle points move into complex plane and the integral (\ref{2-18})
includes the factor $\exp[-\im(p_s(y))y]$ exponentially depending on $y$. Hence, $y=y_f$ corresponds
to the edge of a large amplitude pulse. At the point of transition of two real roots into two complex
ones the equation (\ref{2-19}) has a double root, that is here we have
\begin{equation}\label{2-20}
\frac{d^2\Omega}{dp^2}=0,
\end{equation}
which means that the edge of the pulse corresponds to the extremum of the group velocity.
Just this situation occurs in the case of the dispersion law $\Omega(p)$ of disturbances propagating along a
dark soliton, as one can see in Fig.~1, where $p_{cr}$ corresponds to the minimum (\ref{2-16})
of the group velocity.

It is instructive to illustrate the above consideration by the example of shallow dark solitons
which evolution is described by the Kadomtsev-Petviashvili equation \cite{kp-1970} and the dispersion
law of disturbances was found in explicit form by V.~E.~Zakharov in \cite{zakharov-1975}:
\begin{equation}\label{2-21}
\omega(p) = \frac{\sqrt{2}}{3^{3/4}} p \sqrt{p-p_c},\quad  p_c=\sqrt{3}(1-V),
\end{equation}
where $1-V \ll 1$. From Eq.~(\ref{2-19}) we find the solution for the saddle point
\begin{equation}\label{2-22}
p_s=\frac{1}{3} \left [ 2p_c+\sqrt{3}\frac{y^2}{t^2}+\frac{y}{t}\sqrt{3\frac{y^2}{t^2}-2\sqrt{3}p_c} \right ]
\end{equation}
which is complex for $y^2/t^2<2p_c/\sqrt{3}$. Thus, here the disturbance has the amplitude proportional to
\begin{equation}\label{2-23}
\phi(y, t) \propto \exp \left[ -\frac{y^2}{t} \sqrt{2\sqrt{3}p_c-3\frac{y^2}{t^2}} \right].
\end{equation}
Its edge points propagate with velocities
\begin{equation}\label{2-24}
\frac{y_{\pm}}{t} = \pm \frac{\sqrt{2p_c}}{3^{1/4}} = \pm \sqrt{2(1-V)}
\end{equation}
equal (in absolute value) to the minimum of the group velocity at $p=p_{cr}=\frac{4}{3}p_c$,
\begin{equation}\label{2-25}
V_{cr} = \min \left ( \frac{d\omega}{dp} \right) = \left. \frac{d\omega}{dp} \right|_{p=p_{cr}},
\end{equation}
corresponding to the dispersion law (\ref{2-21}).

\begin{figure}[bt]
\begin{center}
\includegraphics[width=8cm,height=8cm,clip]{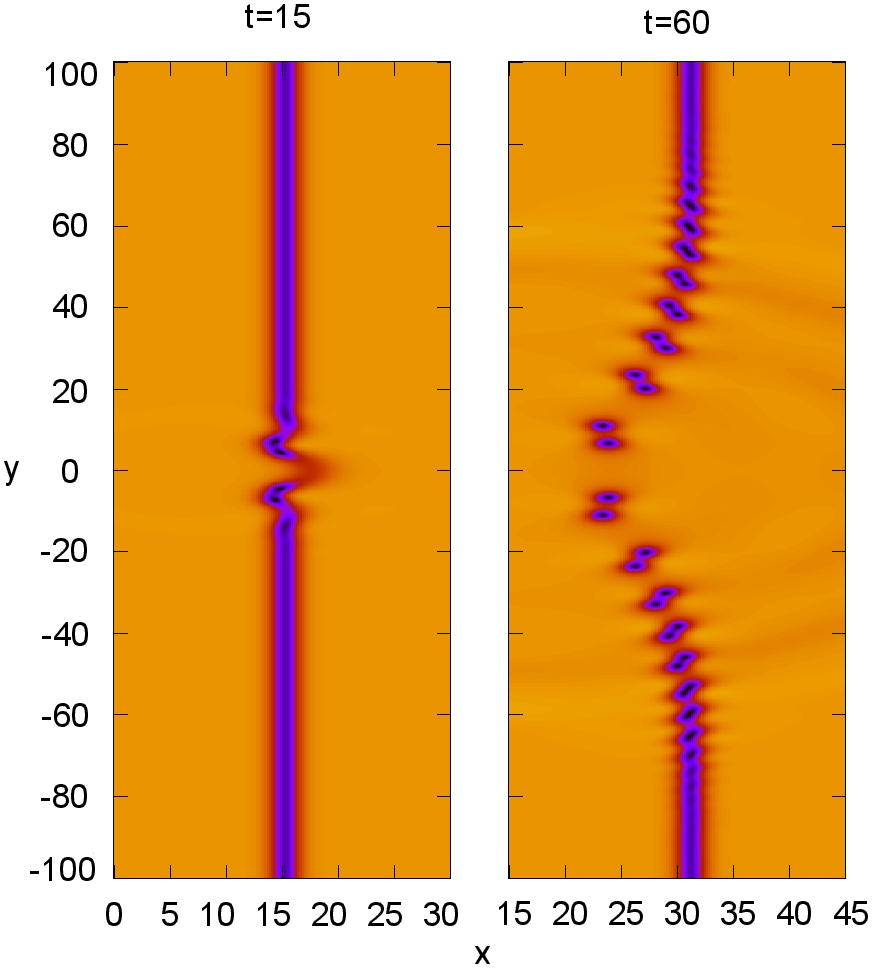}
\caption{Evolution of a localized disturbance with time. A plane soliton slightly disturbed in vicinity of
the point $y=0$ starts its motion at $x=10$ with velocity $V=0.355$. The edges of a disturbed region
expand with velocities equal to $V_f\cong\pm 1.30$.
 }
\end{center}\label{fig2}
\end{figure}
Thus, we can say that for large enough time the edges of the disturbance of a dark soliton propagate
into its undisturbed parts with the universal velocity $V_{cr}$ determined by the parameters of
the soliton rather than the initial form of the disturbance. We have checked this conclusion by
numerical solution of the GP equation (\ref{1-1}) with the initial condition in the form of the
dark soliton (\ref{2-1}) slightly disturbed at point $y=0$. Two stages of evolution are shown in Fig.~2.
Although the disturbed region evolves fast into the nonlinear stage with formation of vortices,
its edge points propagate with velocity $V_f\cong1.30$, which agrees quite well with the critical velocity
$V_{cr}=1.27$ calculated according to the linear theory by Eq.~(\ref{2-16}).

\section{Growth rate of the length of oblique solitons}

\begin{figure}[bt]
\begin{center}
\includegraphics[width=10cm,height=8cm,clip]{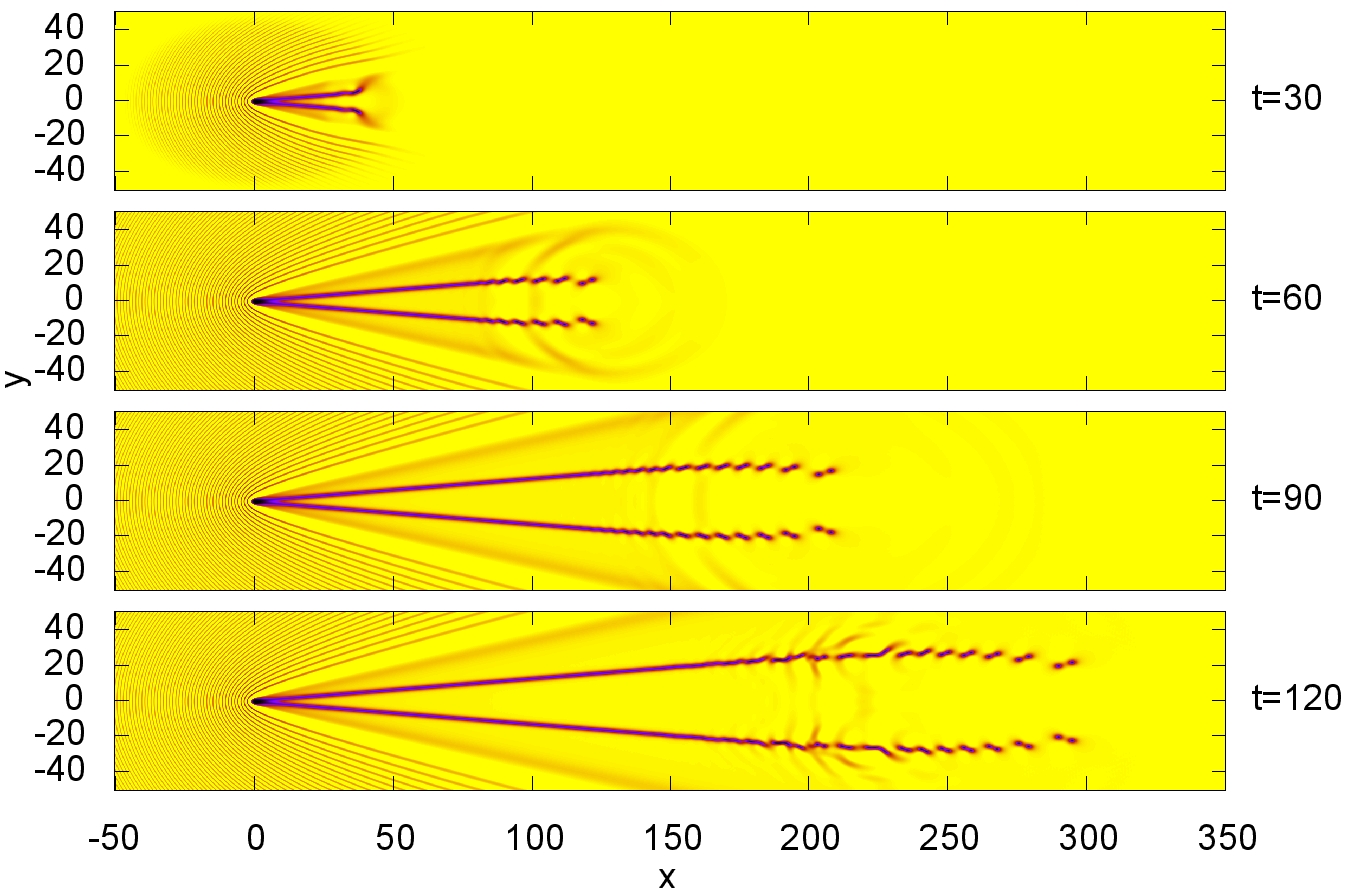}
\caption{Formation of oblique solitons by the flow of the condensate past an obstacle located at the
origin of the coordinate system. To avoid generation of the switching wave at $t=0$, the obstacle
potential was turned on gradually during fist 20 units of time. The Mach number of the flow is equal
to $M=3$ and the oblique solitons correspond to the normal component of the flow velocity $V_{\perp}=0.355$
equal to velocity of the soliton shown in Fig.~2. The length of the oblique solitons grows up with
the rate $dL/dt\cong 1.69$ in agreement with the analytical prediction $dL/dt=1.71$
calculated according to Eq.~(\ref{2-26}).
 }
\end{center}\label{fig3}
\end{figure}
As was predicted theoretically in \cite{egk-2006} for atomic condensate and was confirmed experimentally in
\cite{amo-2011} for polariton condensate, a supersonic flow of condensate past an obstacle with size
about one healing length generates a pair of oblique solitons behind the obstacle provided the incident
flow has large enough Mach number $M>M_{cr}$ (see (\ref{1-3})). If the size of the obstacle is increased,
then two symmetrically situated fans of dark solitons are generated \cite{ek-2006,slender}.
After switching on the flow velocity (or after the obstacle is put into motion), the length $L$
of oblique solitons increases with time linearly and, under supposition that the edge of the soliton can be
considered as the edge of the disturbance propagating into the undisturbed region, the above theory leads to
the conclusion that
\begin{equation}\label{2-26}
\frac{dL}{dt} = V_{\parallel} - V_{cr},
\end{equation}
where
\begin{equation}\label{2-27}
V_{\parallel} = M \cos \theta
\end{equation}
is a projection of the flow velocity on the dark soliton direction (hence the soliton velocity $V$ is
canceled by the normal component $V_{\perp} = M \cos \theta$ of the flow velocity and the oblique
soliton is a stationary object in the laboratory frame of reference). The velocity $V_{cr}$ is
defined by Eq.~(\ref{2-16}) and represents the edge velocity of the disturbance which destroys
the soliton into vortices. For $V_{\parallel} > V_{cr}$ this disturbance is convected away from
the obstacle and the oblique soliton increases its length $L$ with the constant rate (\ref{2-26}).

We have checked this prediction by numerical simulations similar to ones performed in \cite{egk-2006}
but for longer time. Several stages of evolution are shown in Fig.~3. As we found, the oblique
solitons do not stop their growth even after long time of evolution in agrement with the theory
\cite{kp-2008} and the growth rate agrees very well with the analytical prediction (\ref{2-26}).
In the case of the chosen parameters we have $V_{\perp}=0.355$,
the component of the flow velocity along the oblique soliton is equal to
$V_{\parallel}=M\cos\theta=(M^2-V_{\perp}^2)^{1/2}=2.98$, and after subtraction of the minimal
group velocity $V_{cr}=1.27$ corresponding to $p_{cr}=1.08$ we get $dL/dt=1.71$ practically coinciding with the
numerical estimate $dL/dt\cong 1.69$ obtained by fitting to the results
extracted from Fig.~3.

\section{Conclusion}

In this letter we relate the critical velocity of transition from absolute instability
of dark solitons to their convective instability with the velocity of expansion of an unstable
wave packet into undisturbed region. A simple estimate shows that the expansion velocity
is equal to the minimal value of the group velocity of perturbations propagating along the soliton.
This interpretation of the absolute/convective instability transition is physically clear
and leads to experimentally verifiable predictions. In particular, the growth rate of the length of
oblique solitons generated by the flow of BEC past small obstacles can be easily calculated.
The developed theory can also be applied to description of stability of oblique solitons
generated by the flow past extended obstacles. For example, as was observed in \cite{hi-2009},
oblique dispersive shocks generated by the flow of BEC past a concave corner start to decay
at some moment of time with formation of a cloud of vortices.
The character of this instability depends on the projection of the flow velocity on the
inclined side of the corner: if it is greater than the minimal group velocity of perturbations,
then the instability is absolute and the dispersive shock wave will eventually be completely
destroyed; otherwise the instability is convective and the cloud of vortices is convected
away by the flow of BEC as long as this cloud does not interact with the
boundary of the corner and, as a result of this interaction, the convective 
flow along the corner is not destroyed.

\subsection*{Acknowledgments}

We are grateful to G.~A.~El, Yu.~G.~Gladush, and M.~A.~Hoefer for stimulating discussions.
This work was supported by RFBR (grant 09-02-00499).

\end{document}